\documentclass{article}
\usepackage{smc}
\usepackage{times}
\usepackage{ifpdf}
\usepackage[english]{babel}

\usepackage{amsmath,cite,url}
\usepackage{color}

\usepackage{tikz}
\usepackage{multirow}
\usepackage{subcaption}
\usepackage{amssymb}
\usepackage[mathscr]{eucal}
\usepackage{amsmath,amssymb,amsfonts}
\usepackage{algorithmic}
\usepackage{graphicx}
\usepackage{textcomp}
\usepackage{xcolor}
\usepackage{wrapfig}

\def\papertitle{BARWISE COMPRESSION SCHEMES \\ FOR AUDIO-BASED MUSIC STRUCTURE ANALYSIS}
\def\firstauthor{Axel Marmoret}
\def\secondauthor{Jérémy E. Cohen}
\def\thirdauthor{Frédéric Bimbot}

\newcommand{\drawmatr}[6]{
    \draw[black,fill=#6] (#1,#2,#3) -- ++(#4,0,0) -- ++(0,-#5,0) -- ++(-#4,0,0) -- cycle;
}

\newcommand{\drawtensinslices}[8]{
  \foreach \x in {#8,...,0}
    \drawmatr{#1}{#2}{#3-\x*#6/#8}{#4}{#5}{#7};
}

\newif\ifpdf
\ifx\pdfoutput\relax
\else
   \ifcase\pdfoutput
      \pdffalse
   \else
      \pdftrue
\fi

\ifpdf 
  \usepackage[pdftex,
    pdftitle={\papertitle},
    pdfauthor={\firstauthor, \secondauthor, \thirdauthor},
    bookmarksnumbered, 
    pdfstartview=XYZ 
   ]{hyperref}

  \usepackage[figure,table]{hypcap}

\else 
  \usepackage[dvips,
    bookmarksnumbered, 
    pdfstartview=XYZ 
  ]{hyperref}  

  \usepackage[dvips]{epsfig,graphicx}
  \graphicspath{{./figures/}}
  \DeclareGraphicsExtensions{.eps}

  \usepackage[figure,table]{hypcap}
\fi

\hypersetup{
    colorlinks,%
    citecolor=black,%
    filecolor=black,%
    linkcolor=black,%
    urlcolor=black
}

\title{\papertitle}

\def\x{{\mathbf x}}

\DeclareMathOperator*{\argmin}{arg\,min}

 \twoauthors
   {\firstauthor \qquad \thirdauthor} {Univ Rennes, Inria, CNRS, IRISA, France. \\ %
     {\tt \href{mailto:axel.marmoret@irisa.fr}{firstname.name@irisa.fr}}}
   {\secondauthor} {Univ Lyon, INSA-Lyon, UCBL, \\ UJM-Saint Etienne, CNRS, Inserm, \\
CREATIS UMR5220, U1206, France  \\ %
     {\tt \href{mailto:jeremy.cohen@cnrs.fr}{jeremy.cohen@cnrs.fr}}}
\begin{document}
%
\capstartfalse
\maketitle
\capstarttrue
\sloppy
\begin{abstract}
Music Structure Analysis (MSA) consists in segmenting a music piece in several distinct sections. We approach MSA within a compression framework, under the hypothesis that the structure is more easily revealed by a simplified representation of the original content of the song. More specifically, under the hypothesis that MSA is correlated with similarities occurring at the bar scale, this article introduces the use of linear and non-linear compression schemes on barwise audio signals. Compressed representations capture the most salient components of the different bars in the song and are then used to infer the song structure using a dynamic programming algorithm. This work explores both low-rank approximation models such as Principal Component Analysis or Nonnegative Matrix Factorization and ``piece-specific'' Auto-Encoding Neural Networks, with the objective to learn latent representations specific to a given song. Such approaches do not rely on supervision nor annotations, which are well-known to be tedious to collect and possibly ambiguous in MSA description. In our experiments, several unsupervised compression schemes achieve a level of performance comparable to that of state-of-the-art supervised methods (for 3s tolerance) on the RWC-Pop dataset, showcasing the importance of the barwise compression processing for MSA.

\end{abstract}

\section{Introduction}\label{sec:introduction}
Music Structure Analysis (MSA) consists in subdividing a music piece in several distinct parts, which represent a mid-level description of a song. Structure is an important part of music, and could be the main difference between music and random noise~\cite{paulusaudio}. In that sense, MSA has been extensively studied in Music Information Retrieval (MIR), see~\cite{paulusaudio, nieto2020segmentationreview} for overviews. Practically, in the era of computational analysis of music with Machine Learning algorithms, structure could be an important mid-level feature for tasks such as cover detection, genre recognition, music summarization, for better recommendation systems, and many other tasks. Segmentation is usually based on criteria such as homogeneity, novelty, repetition and regularity~\cite{nieto2020segmentationreview}. When performed algorithmically, MSA often relies on similarity criteria within passages of a song summarized in an autosimilarity matrix~\cite{foote2000automatic,serra2014unsupervised, mcfee2014analyzing, nieto2013convex,theodorakopoulos2013unsupervised,mccallum2019unsupervised, marmoret2020uncovering, wang2021supervised, salamon2021deep}, in which each coefficient represents an estimation of the similarity between two musical fragments. Related work mainly splits in two categories: blind (unsupervised) and learning-based (supervised) techniques. Blind segmentation techniques, such as this work, do not use training datasets.

Similarity between two frames can be obtained from the feature representation of the signal, such as the Short-Time Fourier Transform (STFT) of the song~\cite{foote2000automatic}. Boundaries can then be detected as points of strong dissimilarity between the near past and future of a given instant, as in~\cite{foote2000automatic}. Still, recent works have aimed at designing new representations of the original music, able to capture the similarity between two frames while maintaining a high level of dissimilarity between dissimilar frames, either in a blind~\cite{nieto2020segmentationreview,serra2014unsupervised, mcfee2014analyzing, nieto2013convex,theodorakopoulos2013unsupervised, mccallum2019unsupervised, marmoret2020uncovering} or in a supervised~\cite{wang2021supervised, salamon2021deep} fashion. This generally consists in projecting the data in a new feature space and computing the similarity in the feature space.
In particular, McFee and Ellis made use of spectral clustering as a segmentation method by interpreting the autosimilarity as a graph and clustering principally connected vertices as segments~\cite{mcfee2014analyzing}. This work performed best among blind segmentation techniques in the last structural segmentation MIREX campaign in 2016~\cite{MIREXsite}. Recently, we used tensor decomposition (Nonnegative Tucker Decomposition, NTD) as a way to describe music as barwise patterns, which then served as features for the computation of the autosimilarity matrix~\cite{marmoret2020uncovering}.

In most former works, the similarity is expressed between beatwise aligned features, as in~\cite{mcfee2014analyzing, nieto2013convex}. Instead, the present work considers that repetitions are more prone to happen at the barscale, and hence focuses on barwise aligned features, as in~\cite{marmoret2020uncovering}. A comparison between both beatwise and barwise aligned features has been made in~\cite{wang2021supervised}, without one singling out. In this work, we first explore low-rank approximation methods to perform barwise dimensionality reduction using Principal Component Analysis (PCA) and Nonnegative Matrix Factorization (NMF), and then design ``piece-specific'' Auto-Encoding Neural Networks (AE). While PCA and NMF are standard methods nowadays, the work presented in this article is the first one, as far as we know, to consider them for barwise dimensionality reduction.

This work extends the concept of barwise compression by introducing a song-dependent autoencoder, \textit{i.e.} an AE which is specifically trained to compress a given song. This technique is called ``Single-Song Auto-Encoding'' and the latent representation of each bar in the AE is used as a sequence of features for segmenting the song. Indeed, recently, Deep Neural Networks (DNN) have lead to some high level of excitement in MIR research, and notably in MSA~\cite{grill2015music, mccallum2019unsupervised}. In general, DNN approaches rely on large databases which make it possible to learn a large number of parameters, which in turn yields better performance than previously established machine learning approaches. In particular, to the best of the authors' knowledge, the current state-of-the-art approach for the structural segmentation of Popular music 
is a supervised CNN developed by Grill and Schlüter~\cite{grill2015music}. 
This is the consequence of the ability of DNNs to learn complex nonlinear mappings through which musical objects can be expected to be better separated~\cite{humphrey2013feature}. Hence, while DNNs generally learn ``deep'' features stemming from multiple examples in a training phase, and then evaluate the potential of learned features~\cite{agrawal2021learning}, our single-song AE approach focuses on the different events within a single song and tries to learn nonlinear latent representations, used to infer the structure.

To study the relevance of the barwise compression approach, this work evaluates different dimensionality reduction techniques 
on the RWC-Pop dataset~\cite{goto2002rwc} in their audio form using various time-frequency features. Segmentation results on this database show levels of performance which are outperforming the current blind state-of-the-art~\cite{foote2000automatic, mcfee2014analyzing, marmoret2020uncovering}, and our best performing model outperforms the supervised state-of-the-art~\cite{grill2015music} with 3-seconds tolerance.

The rest of the article is structured as follows: Section~\ref{sec:barwise_music_compression} details the motivations and approaches for barwise music compression. Section~\ref{sec:compression_schemes} presents the various compression schemes used in this work. Section~\ref{sec:structural_segmentation} presents the segmentation process. Section~\ref{sec:experiments} reports on the experimental results.

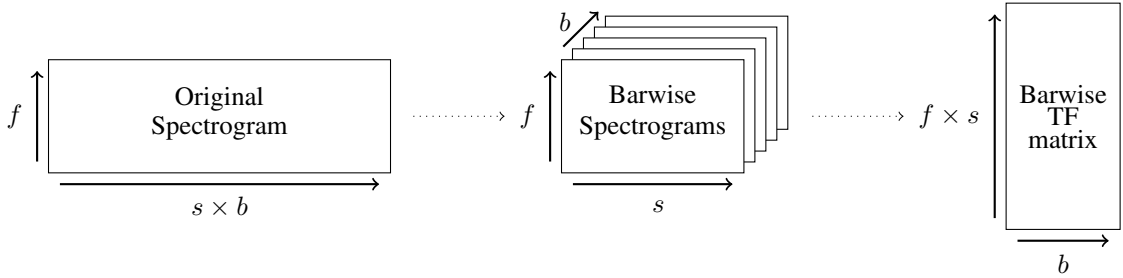
\begin{figure*}[ht]
    \centering
    \begin{tikzpicture}[scale=1.5]
    \drawmatr{0}{0}{0}{3}{1}{white}
    \node at (1.5,-0.35,0) {Original};
    \node at (1.5,-0.6,0) {Spectrogram};
    \draw[->,thick,black] (-0.1,-0.9,0)--(-0.1,-0.1,0);
    \node at (-0.3,-0.5,0) {$f$};
    \draw[->,thick,black] (0.1,-1.1,0)--(2.9,-1.1,0);
    \node at (1.5,-1.3,0) {$s \times b$};

    \draw[->,dotted,black] (3.2,-0.5,0)--(4,-0.5,0);

    \drawtensinslices{4.5}{0}{0}{1.6}{1}{1}{white}{4}
    \draw[->,thick,black] (4.4,-0.9,0)--(4.4,-0.1,0);
    \node at (4.2,-0.5,0) {$f$};
    \draw[->,thick,black] (4.6,-1.1,0)--(6,-1.1,0);
    \node at (5.35,-1.3,0) {$s$};
    \draw[->,thick,black] (4.4,0,-0.3)--(4.4,0,-1.1);
    \node at (4.3,0.1,-0.6) {$b$};
    \node at (5.3,-0.3,0) {Barwise};
    \node at (5.3,-0.6,0) {Spectrograms};

    \draw[->,dotted,black] (6.7,-0.5,0)--(7.5,-0.5,0);

    \drawmatr{8.4}{0.5}{0}{1}{2}{white}
    \draw[->,thick,black] (8.3,-1.4,0)--(8.3,0.4,0);
    \node at (7.9,-0.5,0) {$f \times s$};
    \draw[->,thick,black] (8.5,-1.6,0)--(9.3,-1.6,0);
    \node at (8.9,-1.8,0) {$b$};
    \node at (8.9,-0.3,0) {Barwise};
    \node at (8.9,-0.5,0) {TF};
    \node at (8.9,-0.7,0) {matrix};

    \end{tikzpicture}
    \caption{Barwise processing of an input spectrogram, resulting in a barwise TF matrix.}
    \label{fig:barwise_process}
\end{figure*}

\section{Barwise Music Compression}\label{sec:barwise_music_compression}

\subsection{Motivations}
The underlying idea of this work is that music structure can be related to compression of information. Indeed, a common view of music structure is to consider structural segments as internally coherent passages, and automatic retrieval techniques generally focus on finding them by maximizing homogeneity and repetition and/or by setting boundaries between dissimilar segments, as points of high novelty~\cite{nieto2020segmentationreview}. In the context of compressed representations, each passage is transformed in a vector of small dimension, compelling this representation to summarize the original content. From this angle, similar passages are expected to be represented by similar representations, as they share underlying properties (such as coherence and redundancy), while dissimilar passages are bound to create strong discrepancies at their boundaries. Thus, we expect that compressed representations will enhance the original structure while reducing incidental signal-wise properties which do not contribute to the structure. To the best of our knowledge, this work, our former work~\cite{marmoret2020uncovering}, and recent work by Wang \textit{et al.}~\cite{wang2021supervised} are the only barwise compression schemes with application to structural segmentation of music.

Conversely to fixed-size frame analysis, barwise computation guarantees that the information contained in each frame does not depend on the tempo, but on the metrical positions, which is a more abstract musical notion to describe time.
As a consequence, comparing bars is more reliable than comparing frames of arbitrarily fixed size as it allows to cope with small variations of tempo. In addition, pop music (\textit{i.e.} our case study) is generally quite regular at the bar level: repetitions occur at the bar scale and motivic patterns tend to develop within a limited number of bars, suggesting that frontiers mainly sit between bars. To support this claim, Table~\ref{table:results_sota} of Section~\ref{sec:experiments} presents segmentation results when realigning the annotation on the closest downbeat.

Accordingly, the proposed method relies on a consistent bar division of music. It also requires a powerful tool to detect bars, as otherwise errors could propagate and affect the performance. Results reported in~\cite{marmoret2020uncovering} and in Table~\ref{table:results_sota} tend to show that the madmom toolbox~\cite{madmom} is efficient in that respect on the RWC-Pop dataset. Nonetheless, barwise processing may hinder the retrieval of boundaries delimiting changes of tempo, as discussed in~\cite{vatolkin2021evolutionary}. It also relies on some consistent bar division of music, which is generally the case for contemporary western music, but is not a universal rule, and/or may turn out to be ambiguous.

\subsection{Barwise Music Processing}
Following our former work in~\cite{marmoret2020uncovering}, we process music as barwise spectrograms, with a fixed number of frames per bar. Practically, spectrograms are computed using librosa~\cite{mcfee2015librosa} with a low hop length of 32 frames at a sampling rate of 44.1kHz, and downbeats are estimated with the madmom toolbox~\cite{madmom}. This allows us to split the original spectrogram in $b$ barwise spectrograms ($b$ being the number of bars in this song) each containing $n_b$ frames. As bars can be of different lengths (because of tempi differences or small inconsistencies/imperfections in the performance), different bars can contain different numbers $n_b$ of frames. Thus, we define an integer $s$, the subdivision parameter, which is the desired number of frames in each bar.

Starting from the subdivision $s$, and from indexes $f_s$ and $f_e$, respectively the indexes of the closest frames to the downbeats starting and ending the bar, we select all frames $\Big\{\frac{f_s + k \times (f_e - f_s)}{s}, 0 \leq k < s, k \in \mathbb{N}\Big\}$, \textit{i.e.} equally-spaced frames in the bar to fit the chosen subdivision. This indeed results in barwise spectrograms containing exactly $s$ frames. Other techniques could be applied to reduce the number of frames (for example averaging the content of several frames instead of choosing one), but we did not pursue that lead. We chose $s = 96$ as in~\cite{marmoret2020uncovering}.

This technique result in $b$ barwise Time-Frequency spectrograms, of size $f\times s$, with $f$ the number of components in the chosen feature, as presented in section~\ref{sec:features}. Compression is generally performed on matrices (as PCA and NMF for instance). In that sense, by vectorizing each of the previous barwise spectrograms, we introduce the ``barwise TF'' representation, consisting in a matrix of size $b \times (f \times s)$. 
The aforementioned process is described in Figure~\ref{fig:barwise_process}. Note that we chose to vectorize the time-frequency features, thus discarding the dependencies between these two dimensions.


\subsection{Features}\label{sec:features}
Music is represented with different features throughout our experiments, focusing on different aspects of music such as harmony or timbre. These features are presented hereafter.

\textbf{Chromagram} A chromagram represents the time-frequency aspect of music as sequences of 12-row vectors, corresponding to the 12 semi-tones of the classical western music chromatic scale (C, C\#, ..., B), which is largely used in Pop music. Hence, $f = 12$, and each row represents the weight of a semi-tone (and its octave counterparts) at a particular instant.

\textbf{Mel spectrogram} A Mel spectrogram corresponds to the STFT representation of a song, whose frequency bins are recast in the Mel scale. These bands account for the exponential spread of frequencies throughout the octaves. Mel spectrograms are dimensioned following the work of~\cite{grill2015music} (80 filters, from 80Hz to 16kHz), hence $f = 80$. STFT are computed as power spectrograms.

\textbf{Log Mel spectrogram (LMS)} A Log Mel spectrogram (abbreviated ``LMS'') corresponds to the logarithmic values of the precedent spectrogram, hence $f$ is still equal to $80$. This representation accounts for the exponential decay of power with frequency observed in STFT.

\textbf{Nonnegative Log Mel spectrogram (NNLMS)} Nonnegative Log Mel spectrogram (abbreviated ``NNLMS'') is a nonnegative version of the precedent LMS. This representation is motivated by the fact that some compression algorithms are nonnegative (in particular, NMF), and thus need nonnegative inputs. NNLMS are computed as $\log(Mel + 1)$ where $Mel$ represent the coefficients of the Mel spectrogram, which are nonnegative, and $\log$ the elementwise logarithm.

\textbf{MFCC spectrogram} Mel-Frequency Cepstral Coefficients (MFCCs) are timbre-related coefficients, obtained by a discrete cosine transform of a Log Mel spectrogram\footnote{Note that we use the default librosa settings for MFCC, hence the Log Mel spectrogram used for MFCC is not the same as for the LMS.}. This spectrogram contains $f = 32$ coefficients, following~\cite{mcfee2014learning}.

\section{Compression Schemes}\label{sec:compression_schemes}
\subsection{Low-Rank Approximations}
\label{sec:lra}
Given a collection of input vectors $\{x_i\}_{i=1, \ldots, b}\in\mathbb{R}^{n}$ stored in a matrix $X$, low-rank approximations search for projections in a low-dimensional subspace 
which approximates the input vectors. We study two different low-ranks approximations methods, namely PCA and NMF.

PCA is maybe the most well-known and used low-rank approximation technique. It can be seen as a singular value decomposition applied on the debiased matrix $X-\mu$, where $\mu$ is the mean of the columns of $X$. In short, we obtain an approximation $X-\mu \approx WH$ where $W\in\mathbb{R}^{n\times d_c}$ is orthogonal, and $H\in\mathbb{R}^{d_c\times b}$ is the product of a diagonal and an orthogonal matrix. The rank of the approximation $d_c$ is chosen by the user, and should be smaller than both dimensions.

NMF is similar to PCA but does not impose orthogonality, nor does it unbias the data. Instead, it builds a cone containing the data, which extreme rays are the columns of $W$. To achieve this, nonnegativity is required elementwise on both $W$ and $H$. The low-rank approximation is in fact the solution of an optimization problem, here:
\begin{equation}
  \underset{W\geq 0, H\geq 0}{\argmin} \|X - WH\|_F^2
\end{equation} using the Frobenius norm as a loss function. NMF typically yields more interpretable features than PCA, but it is much harder to compute, and requires that the data is mostly nonnegative~\cite{GillisBookNMF}. In both cases, $H$ is the barwise compressed representation.

\subsection{Autoencoders}
\label{sec:ae_generalities}
Autoencoders are neural networks, which, by design, perform unsupervised dimensionality reduction. Throughout the years, AE have received increasing interest, notably due to their ability to extract salient latent representations without the need of large amount of annotations. Recently, AE also showed great results as a generation tool~\cite{engel2017neural, roche2018autoencoders}. Still, as presented in~\cite{roche2018autoencoders}, PCA and AE are competitive as compression schemes, and PCA even leads to lower reconstruction error when AE are too ``simple'' (in particular when networks are linear or ``shallow'', \textit{i.e.} with only a few layers).

Practically, given a generic entry $x \in \mathbb{R}^n$, an AE learns a nonlinear function $f$ with parameters $\theta$ (weights and biases of the network) such that $\hat{x} = f(x, \theta) \in \mathbb{R}^n$ reconstructs $x$ as faithfully as possible. This is achieved by minimizing a given loss function such as the Mean Square Error (MSE) between $x$ and $\hat{x}$ w.r.t. parameters $\theta$.
\begin{equation}
  \underset{\theta}{\argmin} \text{ MSE}(x, \hat{x}) = \frac{1}{n} \sum_{i = 0}^{n - 1} (x_i - \hat{x}_i)^2~.
\end{equation}
Other metrics can be used, such as the Kullback-Leibler divergence, but we restrict this work to MSE.

An autoencoder is divided into two parts: an encoder, which compresses the input $x \in \mathbb{R}^n$ into a latent representation $z \in \mathbb{R}^{d_c}$ of smaller dimension (generally, $d_c \ll n$), and a decoder, which reconstructs $\hat{x}$ from $z$. While Autoencoders generally learn a common latent representation for an entire dataset, our technique optimizes a network for each song. This framework is called ``single-song autoencoding''.

In this work, layers are of two types: fully-connected, or convolutional. Convolutional layers lead to impressive results in image processing due to their ability to discover local correlations (such as lines or edges), which turn to higher-order features with the depth of the network~\cite{lecun1998gradient}~\cite[Ch.~9]{Goodfellow2016}. While local correlations are less obvious in spectrogram processing~\cite{peeters2021deep}, Convolutional Neural Networks (CNN) still perform well in MIR tasks, such as MSA~\cite{grill2015music}.

The tested AE is hence a CNN. We use the nowadays quite standard Rectified Linear Unit (ReLU) function as the activation function, except in the latent layer because it could lead to null latent variables. The encoder is composed of five hidden layers: 2 convolutional/max-pooling blocks, followed by a fully-connected layer, controlling the size $d_c$ of the latent space. Convolutional kernels are of size 3x3, and the pooling is of size 2x2. Convolutional layers define respectively 4 and 16 feature maps.

\begin{figure}[ht]
 \centerline{\includegraphics[width=\columnwidth]{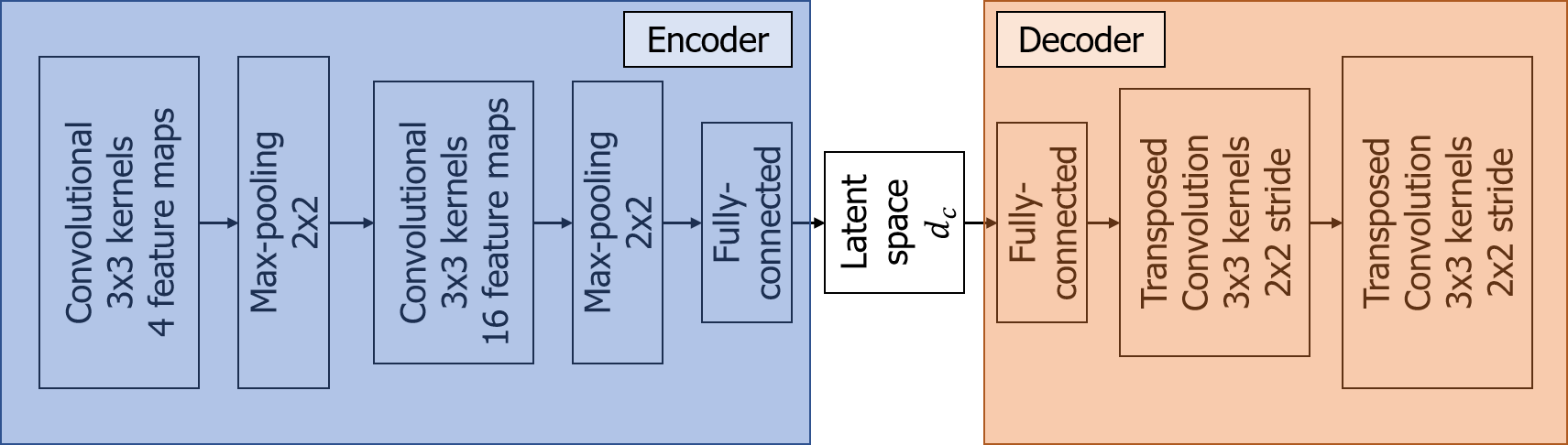}}
 \setlength{\abovecaptionskip}{5pt}
 \caption{Architecture of the autoencoder}
 \label{fig:convolutional}
\end{figure}

The decoder is composed of 3 hidden layers: a fully-connected layer (inverse of the previous one) and 2 ``transposed convolutional'' layers of size 3x3 and stride 2x2. A transposed convolution is similar to the convolution operation taken in the backward pass: an operation which takes one scalar as input and returns several scalars as output~\cite[Ch.~4]{dumoulin2016guide}. Hence, it is well suited to inverse the convolution process. This network is represented in Figure~\ref{fig:convolutional}.

\subsection{Dimensions of Compression}
Selecting the dimension of the compression $d_c$, \textit{i.e.} the number of components to use for compressing the input data, is not trivial. In particular, it is well-known that selecting the number of components for NMF is a complex problem, generally leading to manual tuning or dedicated heuristics~\cite{GillisBookNMF, nebgen2021neural}. We also observed a great influence of the size of the latent space of the AE on the segmentation results, without a clear candidate singling out. In that sense, we will compare values for $d_c \in \{8, 16, 24, 32, 40\}$. Even if such heuristics are standard for PCA (such as the elbow method), as no obvious candidate heuristic stood out relatively to the quality of segmentation, and for fair comparison with other techniques, PCA will be tested with the same $d_c$ values. A clever dimension selection method could be studied in future work.

\section{Structural segmentation}\label{sec:structural_segmentation}
\subsection{General Principle}
\begin{figure*}[ht]
\vspace{-6pt}
 \includegraphics[width=2\columnwidth]{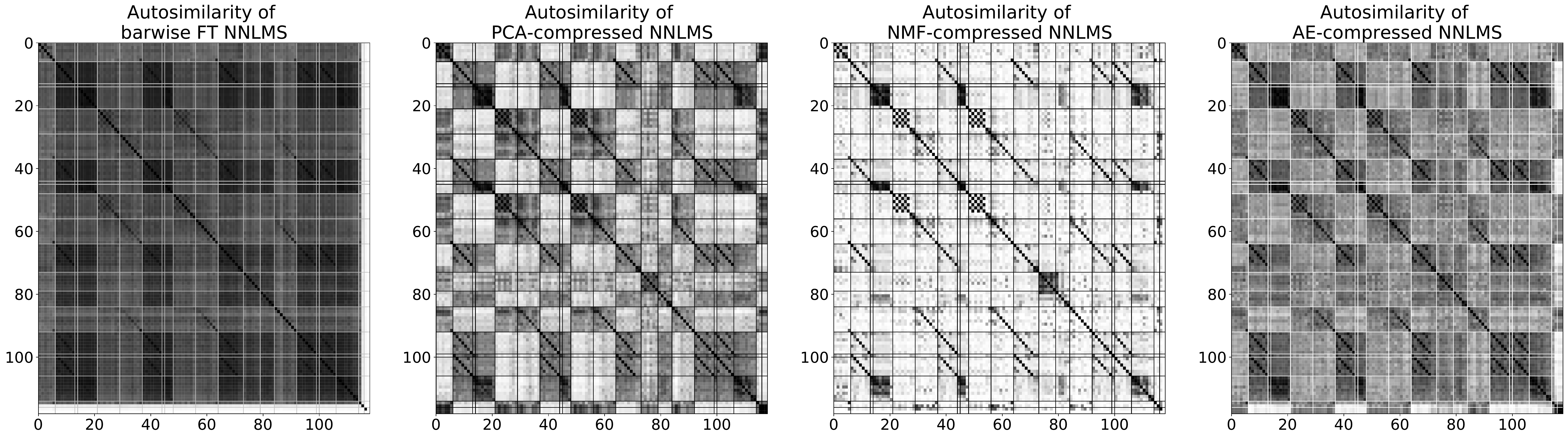}
\setlength{\belowcaptionskip}{-5pt}
 \caption{Barwise autosimilarities for the Nonnegative Log Mel spectrogram (left) and of the compressed representation with $d_c = 24$ (from left to right: PCA, NMF and AE) computed on the song ``Pop01'' from RWC-Pop. Horizontal and vertical lines represent the annotation.}
 \label{fig:autosimilarity}
\end{figure*}

The ability of these compression methods to separate and group bars is evaluated on the MSA task, as presented in~\cite{paulus2010state}. Thereby, for a given song in a given representation, we obtain compressed representations ${z_i}$ for all bars $1 \leq i \leq b$ of the song. These compressed representations are at the heart of the segmentation strategy, via their autosimilarity matrix. Denoting $Z \in \mathbb{R}^{d_c \times b}$ the matrix resulting from the concatenation of all $d_c$-dimensional $z_i$ barwise compressed representations, its autosimilarity is defined as $Z^T Z$, \textit{i.e.} the $b \times b$ matrix of the dot products between every $z_i$. These dot products are then normalized to 1, resulting in a matrix of cosine similarities. Examples of autosimilarities are shown in Figure~\ref{fig:autosimilarity}.

Segmentation is obtained via the dynamic programming algorithm presented in~\cite{marmoret2020uncovering}, and inspired from~\cite{sargent2011regularity}. The principle of dynamic programming is to solve a combinatorial optimization problem by dividing it in several atomic problems, easier to solve, and which solutions can be stitched together to form the global solution. The Viterbi and Dynamic Time Warping (DTW) algorithms are examples of dynamic programming algorithms with a lot of applications in the Audio community. In our context, a ``segmentation cost'' is computed for every segment in the song. These individual segment costs constitute the atomic problems to solve. Then, the optimal segmentation consists in the global maximum of the sum of the segment costs for all the segments in this segmentation.

The cost of each segment is designed in order to favor their homogeneity. As the autosimilarity matrix represent the similarity between pairs of bars in the song, the higher is this similarity, the most similar are the bars. Hence, the cost of a segment is defined as an aggregated value of the similarity between all pairs of bars in this segment. Practically, this is obtained by convolving a kernel with the autosimilarity restricted to this segment. Hence, to compute the cost of each segment in the song, the algorithm applies a sliding convolving kernel on the diagonal of the autosimilarity matrix. This convolving kernel is a square matrix, the size of which is that of the potential segment.

While sharing the principle of a sliding kernel with the work of Foote~\cite{foote2000automatic}, largely described in the literature~\cite{nieto2020segmentationreview}, the proposed kernel focuses on homogeneity rather than novelty. When the diagonal of the autosimilarity is structured in several self-similar blocks, the algorithm frames and partitions these blocks.

\subsection{Technical Details}
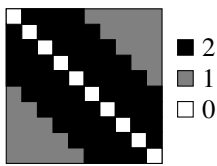
\begin{figure}
  \begin{center}
    \begin{tikzpicture}[scale=2.05]
    \drawmatr{0}{0}{0}{1}{1}{gray}
    \drawmatr{0}{-0.1}{0}{0.1}{0.4}{black}
    \drawmatr{0}{0}{0}{0.1}{0.1}{white}
    \drawmatr{0.1}{-0.1}{0}{0.1}{0.1}{white}
    \drawmatr{0.2}{-0.2}{0}{0.1}{0.1}{white}
    \drawmatr{0.3}{-0.3}{0}{0.1}{0.1}{white}
    \drawmatr{0.4}{-0.4}{0}{0.1}{0.1}{white}
    \drawmatr{0.5}{-0.5}{0}{0.1}{0.1}{white}
    \drawmatr{0.6}{-0.6}{0}{0.1}{0.1}{white}
    \drawmatr{0.7}{-0.7}{0}{0.1}{0.1}{white}
    \drawmatr{0.8}{-0.8}{0}{0.1}{0.1}{white}
    \drawmatr{0.9}{-0.9}{0}{0.1}{0.1}{white}
    \drawmatr{0.1}{-0.2}{0}{0.1}{0.4}{black}
    \drawmatr{0.2}{-0.3}{0}{0.1}{0.4}{black}
    \drawmatr{0.3}{-0.4}{0}{0.1}{0.4}{black}
    \drawmatr{0.4}{-0.5}{0}{0.1}{0.4}{black}
    \drawmatr{0.5}{-0.6}{0}{0.1}{0.4}{black}
    \drawmatr{0.6}{-0.7}{0}{0.1}{0.3}{black}
    \drawmatr{0.7}{-0.8}{0}{0.1}{0.2}{black}
    \drawmatr{0.8}{-0.9}{0}{0.1}{0.1}{black}
    \drawmatr{0.1}{0}{0}{0.4}{0.1}{black}
    \drawmatr{0.2}{-0.1}{0}{0.4}{0.1}{black}
    \drawmatr{0.3}{-0.2}{0}{0.4}{0.1}{black}
    \drawmatr{0.4}{-0.3}{0}{0.4}{0.1}{black}
    \drawmatr{0.5}{-0.4}{0}{0.4}{0.1}{black}
    \drawmatr{0.6}{-0.5}{0}{0.4}{0.1}{black}
    \drawmatr{0.7}{-0.6}{0}{0.3}{0.1}{black}
    \drawmatr{0.8}{-0.7}{0}{0.2}{0.1}{black}
    \drawmatr{0.9}{-0.8}{0}{0.1}{0.1}{black}

    \drawmatr{1.1}{-0.2}{0}{0.1}{0.1}{black}
    \node at (1.3,-0.25) {2};
    \drawmatr{1.1}{-0.4}{0}{0.1}{0.1}{gray}
    \node at (1.3,-0.45) {1};    \drawmatr{1.1}{-0.6}{0}{0.1}{0.1}{white}
    \node at (1.3,-0.65) {0};

\end{tikzpicture}
\end{center}
  \setlength{\abovecaptionskip}{-6pt}%
  \setlength{\belowcaptionskip}{-13pt}%
  \caption{Kernel of size 10}
  \label{fig:kernel}
\end{figure}

The design of the kernel defines how to transform bar similarities into segment homogeneity in the form of a cost. A kernel matrix full of ones would imply that the segment cost is the sum of the similarity values in this segment. The proposed kernel is equal to 0 on the diagonal, 2 on the 4 sub and super diagonals, and 1 everywhere else. The diagonal is equal to zero so as to disregard perfect self-similarity of each bar with itself (normalized to 1). The other elements of this kernel are designed so as to emphasize the short-term similarity in the 4 contiguous bars, and still catch longer-term similarity for long segments. It is presented in Figure~\ref{fig:kernel}. This kernel performs best in comparative experiments\footnote{https://gitlab.inria.fr/amarmore/musicntd/-/blob/v0.2.0/Notebooks/5\%20-\%20Different\%20kernels.ipynb} of~\cite{marmoret2020uncovering}, and, notably, better than a kernel matrix of 1 with a 0 diagonal.

\begin{figure}[ht]
    \centering
    \includegraphics[width=\columnwidth]{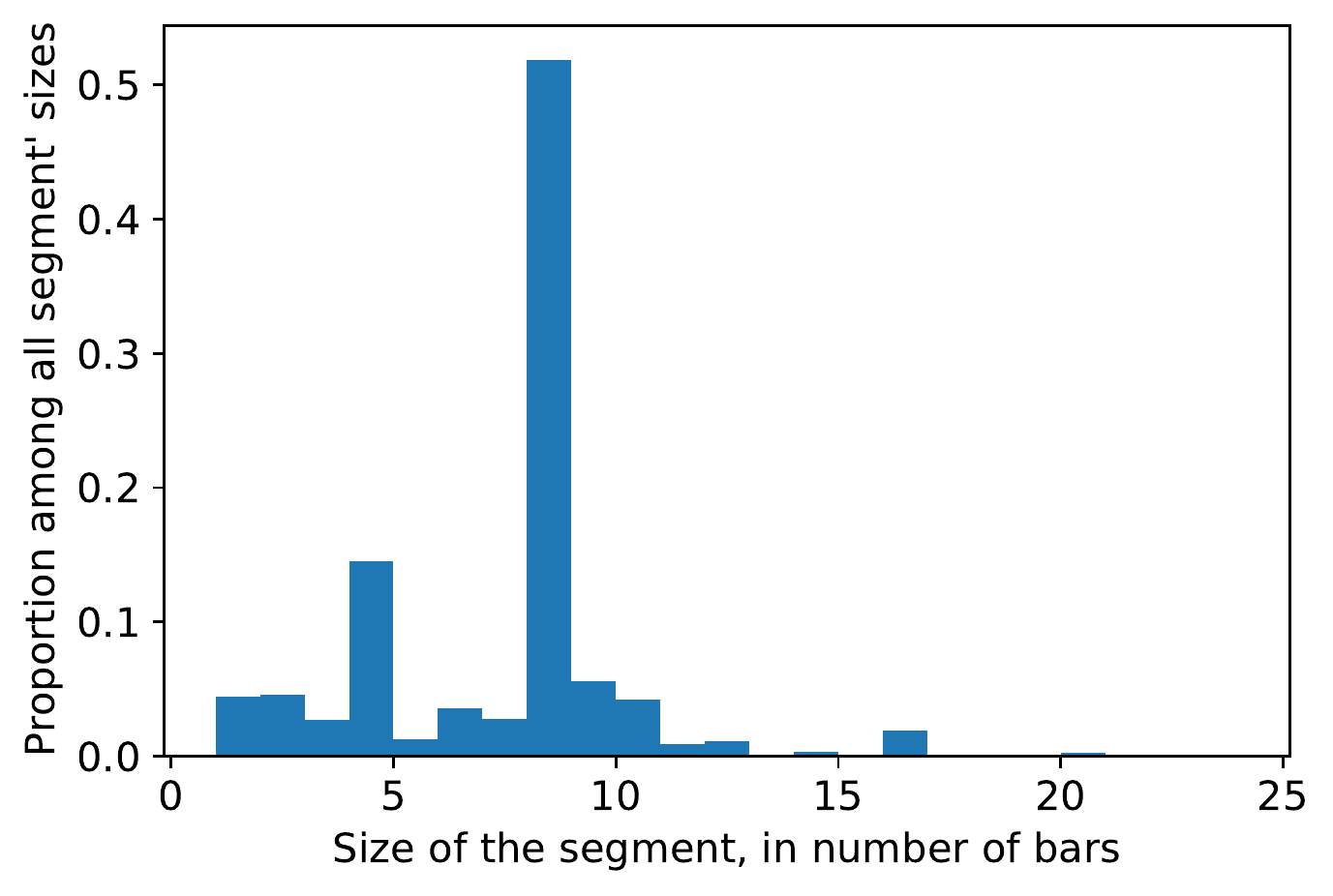}
    \caption{Distribution of segment' sizes in number of bars in MIREX10.}
    \label{fig:histo_seg_size}
\end{figure}

To account for regularity constraints as in~\cite{sargent2011regularity}, the present algorithm adds a regularity penalty $p(n)$ to the plain convolution score, which is a function of the size $n$ (in bars) of the segment. This function $p(n)$ is equal to 0 if $n = 8$, $\frac{1}{4}$ if $n = 4$, $\frac{1}{2}$  if $n \equiv 0 \pmod 2$, and finally 1 otherwise. This penalty function is musically motivated for pop music, as segments are more likely to be of size 4 or 8 bars (especially in RWC-Pop with MIREX10 annotations~\cite{sargent2011regularity}), than of odd bar size, as shown in Figure~\ref{fig:histo_seg_size}.

Finally, for all possible segments $[b_1, b_2[$ in the song, the algorithm computes a score:
\begin{equation}
  s_{b_1,b_2} = \frac{c_{b_1,b_2}}{c_{k8}^{max}} - p(n),
\end{equation}
where $c_{b_1,b_2}$ is the convolution cost, and where $c_{k8}^{max}$ is the highest convolution score on this autosimilarity with a kernel of size 8, used for normalization purposes.

\section{Experiments}\label{sec:experiments}
The proposed segmentation pipeline is studied on the RWC Pop dataset, which consists in 100 Pop songs of high recording quality~\cite{goto2002rwc}, along with the MIREX10 annotations~\cite{bimbot2014semiotic}. We compared the barwise compression schemes described in the present work with three blind methods and a supervised method: Foote's novelty kernel~\cite{foote2000automatic}, Spectral Clustering by McFee and Ellis~\cite{mcfee2014analyzing}, our former NTD method~\cite{marmoret2020uncovering}, and the supervised CNN of Grill and Schlüter~\cite{grill2015music}, the latter being the current state-of-the-art.
Results for~\cite{foote2000automatic} and~\cite{mcfee2014analyzing} were computed with the MSAF toolbox~\cite{nieto2016systematic}, and boundaries were aligned to the closest downbeat for fairer comparison (as in~\cite{marmoret2020uncovering}). Results for~\cite{grill2015music} were taken from the 2015 MIREX campaign~\cite{MIREXsite}.

We focus on boundary retrieval and ignore segment labelling. Boundaries are evaluated using the hit-rate metric, which considers a boundary valid if it is approximately equal to an annotation, within a fixed tolerance window. Consistently with MIREX standards~\cite{MIREXsite}, tolerances are equal to 0.5s and 3s. The hit-rate is then expressed in terms of Precision, Recall, and F-measure, resulting respectively in $P_{0.5}$, $R_{0.5}$, $F_{0.5}$, $P_{3}$, $R_{3}$, $F_{3}$.

\begin{figure*}[ht]
\centering
\begin{subfigure}[ht]{1\columnwidth}
  \includegraphics[width=\columnwidth]{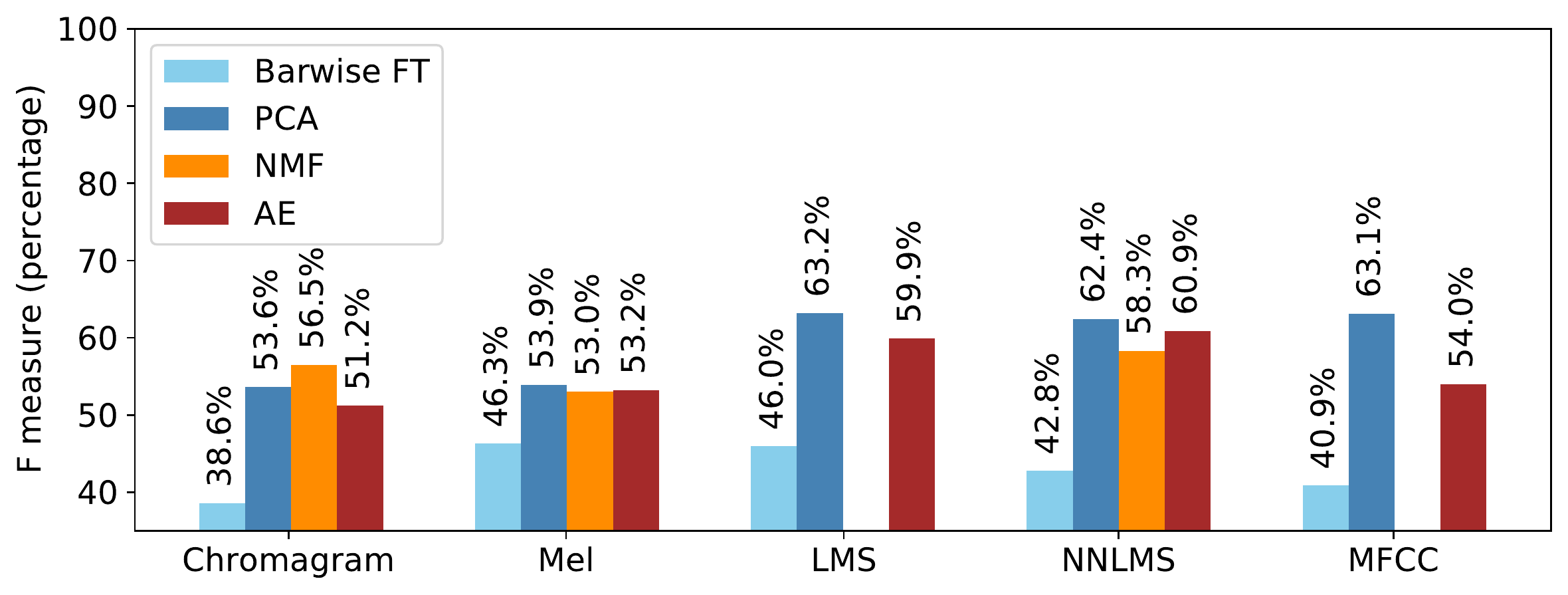}
 \caption{0.5 seconds tolerance results ($F_{0.5}$)}
\end{subfigure}
\quad
\begin{subfigure}[ht]{1\columnwidth}
  \includegraphics[width=\columnwidth]{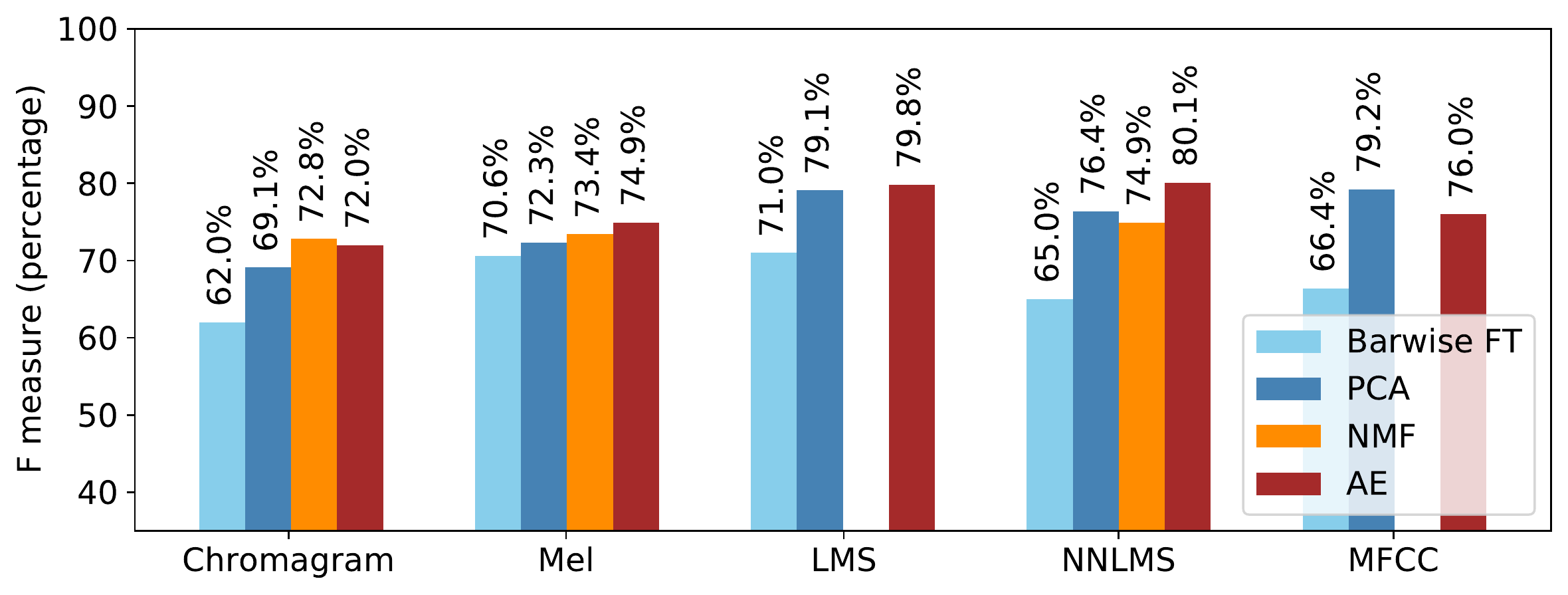}
    \setlength{\abovecaptionskip}{0pt}%
      \caption{3 seconds tolerance results ($F_{3}$)}
  \end{subfigure}
\caption{Segmentation results for the different methods and representation.}
\label{fig:results_all_methods}
\end{figure*}

\begin{table*}[ht]
\caption{Results of best-performing methods and state-of-the-art on the RWC-Pop dataset. (*Results: 2015 MIREX contest~\cite{MIREXsite}.)}
\vspace{-10pt}
\setlength{\abovecaptionskip}{-10pt}%
\setlength{\belowcaptionskip}{-20pt}%
 \begin{center}
\begin{tabular}{ll|c|c|c|c|c|c|}
\cline{3-8} &        & $P_{0.5}$ & $R_{0.5}$ & $F_{0.5}$ & $P_{3}$ & $R_{3}$ & $F_{3}$ \\
\hline

\multicolumn{1}{|l|}{\multirow{3}{*}{\begin{tabular}[c]{@{}l@{}}Compression methods\end{tabular}}} & \multicolumn{1}{l|}{PCA} & \multicolumn{1}{l|}{60.7\%}   &  \multicolumn{1}{l|}{\textbf{66.9\%}}      & \multicolumn{1}{l|}{63.1\%}    & \multicolumn{1}{l|}{76.1\%}  & \multicolumn{1}{l|}{\textbf{84.0\%}}  & \multicolumn{1}{l|}{79.2\%}    \\ \cline{2-8}

\multicolumn{1}{|l|}{} & \multicolumn{1}{l|}{NMF} & \multicolumn{1}{l|}{57.4\%}    & \multicolumn{1}{l|}{60.4\%}      & \multicolumn{1}{l|}{58.3\%}      & \multicolumn{1}{l|}{73.6\%}  & \multicolumn{1}{l|}{77.6\%}  & \multicolumn{1}{l|}{74.9\%}  \\ \cline{2-8}

\multicolumn{1}{|l|}{}  & \multicolumn{1}{l|}{Autoencoder}  & \multicolumn{1}{l|}{60.1\%}    & \multicolumn{1}{l|}{62.6\%}    & \multicolumn{1}{l|}{60.9\%}      & \multicolumn{1}{l|}{78.9\%}  & \multicolumn{1}{l|}{82.5\%}  & \multicolumn{1}{l|}{\textbf{80.1\%}}  \\ \hline \hline

\multicolumn{1}{|l|}{\multirow{4}{*}{\begin{tabular}[c]{@{}l@{}}State-of-the-art\end{tabular}}} & \multicolumn{1}{l|}{Foote~\cite{foote2000automatic}} & \multicolumn{1}{l|}{42.0\%}   &  \multicolumn{1}{l|}{30.0\%}      & \multicolumn{1}{l|}{34.5\%}    & \multicolumn{1}{l|}{67.1\%}  & \multicolumn{1}{l|}{47.7\%}  & \multicolumn{1}{l|}{55.0\%}    \\ \cline{2-8}

\multicolumn{1}{|l|}{} & \multicolumn{1}{l|}{Spectral clustering~\cite{mcfee2014analyzing}} & \multicolumn{1}{l|}{49.2\%}    & \multicolumn{1}{l|}{45.0\%}      & \multicolumn{1}{l|}{45.0\%}      & \multicolumn{1}{l|}{65.5\%}  & \multicolumn{1}{l|}{60.6\%}  & \multicolumn{1}{l|}{60.3\%}  \\ \cline{2-8}
\multicolumn{1}{|l|}{}  & \multicolumn{1}{l|}{NTD~\cite{marmoret2020uncovering}}  & \multicolumn{1}{l|}{58.4\%}    & \multicolumn{1}{l|}{60.7\%}    & \multicolumn{1}{l|}{59.0\%}      & \multicolumn{1}{l|}{72.5\%}  & \multicolumn{1}{l|}{75.3\%}  & \multicolumn{1}{l|}{73.2\%}  \\  \cline{2-8}
\multicolumn{1}{|l|}{} & \multicolumn{1}{l|}{Supervised CNN~\cite{grill2015music}*} &  \multicolumn{1}{l|}{\textbf{80.4\%}}    & \multicolumn{1}{l|}{62.7\%}    & \multicolumn{1}{l|}{\textbf{69.7\%}}    & \multicolumn{1}{l|}{\textbf{91.9\%}}  & \multicolumn{1}{l|}{71.1\%}  & \multicolumn{1}{l|}{79.3\%}  \\ \hline \hline

\multicolumn{2}{|c|}{Aligning annotation on downbeats} &  \multicolumn{1}{l|}{96.5\%}    & \multicolumn{1}{l|}{96.2\%}    & \multicolumn{1}{l|}{96.3\%}    & \multicolumn{1}{l|}{100\%}  & \multicolumn{1}{l|}{99.7\%}  & \multicolumn{1}{l|}{99.9\%}  \\ \hline
\end{tabular}
\end{center}
\label{table:results_sota}
\vspace{-20pt}
\end{table*}

\subsection{Practical Considerations}
PCA is computed with the scikit-learn~\cite{scikitlearn} toolbox, using the ARPACK solver. NMF is computed using the nn\_fac toolbox~\cite{marmoret2020nn_fac}.
The AE is developed with Pytorch 1.8.0~\cite{NEURIPS2019_9015}, trained with the Adam optimizer~\cite{kingma2014adam}, with a learning rate of 0.001, divided by 10 when the loss function reaches a plateau (20 iterations without improvement) until 1e-5. The optimization stops if no progress is made during 100 consecutive epochs, or after a total of 1000 epochs. The network is initialized with the uniform distribution defined in~\cite{he2015delving}, also known as ``kaiming'' initialization. Bars were estimated with madmom~\cite{madmom}, and spectrograms computed with librosa~\cite{mcfee2015librosa} with default settings, unless specified. All segmentation scores were computed with mir\_eval~\cite{raffel2014mireval}.

The entire code for this work is open-source, and contains experimental notebooks for reproducibility\footnote{https://gitlab.inria.fr/amarmore/BarwiseMusicCompression}. Performing 1000 epochs for a song takes between 1.5 minute (for chromagrams) and 5 minutes (for Mel/Log Mel spectrograms) on an Intel\textsuperscript{\textregistered} Core(TM) i7 CPU, NMF takes less than 3 seconds and PCA less than a second.

\subsection{Results}\label{sec:results}
Figure~\ref{fig:results_all_methods} presents results on the RWC-Pop dataset for all the methods and the features. The dimension of compression $d_c$ is chosen as the best performing one for the $F_3$ metric. The Log Mel and Nonnegative Log Mel spectrogram seem to outperform the other features at both tolerances, except for the PCA which obtain similar performances with MFCC. PCA and AE achieve the state-of-the-art level of performance with 3-seconds tolerance, as presented in Table~\ref{table:results_sota}. With 0.5-second tolerance, all techniques obtain lower results than those of the state-of-the-art, but outperform the blind state-of-the-art. Results are consistent with those of~\cite{roche2018autoencoders}, where PCA obtain similar or better reconstruction errors than linear or shallow autoencoders. Still, we believe that the presented AE can be greatly improved.

NMF obtain lower results than the other methods, but, due to the nonnegativity, it is expected that NMF result in a part-based decomposition of the original barwise spectrogram. The part-based property can lead to interpretable factors, as observed in many applications~\cite{lee1999learning, GillisBookNMF}, and is important for pattern uncovering in NTD~\cite{marmoret2020uncovering}. Nonetheless, in this study, we do not have quantitative results to support this claim.


Results seem still improvable with 0.5-second tolerance. A wrong estimation at 0.5s can be due to an incorrect frontier estimation, but also to an incorrect bar estimation. Results when aligning the annotation on downbeats presented in Table~\ref{table:results_sota} indicate that bar estimation seems precise on the RWC Pop dataset.



\section{Conclusion}
This article introduce barwise compression schemes, which appear as competitive schemes for Music Structure Analysis. In our experiments on the RWC-Pop music dataset, segmentation scores obtained with compressed representations outperform the results of the blind state-of-the-art, and our best-performing methods reach the level of performance of the global state-of-the-art at 3-seconds tolerance, while being unsupervised. Still, this work focus on the RWC-Pop dataset, and would benefit from further investigations on different musical contexts. In particular, it would be interesting to study the barwise compression schemes in condition where bars are more ambiguous and/or less regular (polyrhythms/polymeters, changing meters, etc) and on more erratic structures (\textit{i.e.} segments less concentrated around the sizes of 4 and 8 bars).

Future work will focus on improving the proposed paradigm, for instance with nonnlinear compression methods such as kernel methods, or by improving the autoencoder in using strategies such as transfer learning from a song-independent AE or exploring various network architectures or regularizations. Finally, while the convolutional dynamic programming algorithm is competitive, it is unstable to small variations in autosimilarities and should be made more robust.

Altogether, we believe that these first results pave the way to an interesting paradigm using barwise compression algorithms, and notably single-song AEs, for the description of structural elements in music.

\bibliographystyle{IEEEbib}
\bibliography{biblio}

\end{document}